\documentclass[12pt]{article}
\usepackage{epsfig}
\usepackage{float}
\usepackage{graphicx}
\begin{document}
\rightline{NKU-2019-SF1}
\bigskip


\begin{center}
{\Large \bf Non-linear black holes in  (2+1)-dimensions as  heat engines}
\end{center}
\hspace{0.4 cm}
\begin{center}
Leonardo Balart\footnote{leonardo.balart@ufrontera.cl}\\
{\small \it Departamento de Ciencias F\'{\i}sicas, \\
Facultad de Ingenier\'{\i}a y Ciencias \\ \small Universidad de La Frontera, Casilla 54-D \\
Temuco, Chile.}\\
\vspace{0.3 cm}
Sharmanthie Fernando\footnote{fernando@nku.edu}\\
{\small\it Department of Physics, Geology  \& Engineering Technology}\\
{\small\it Northern Kentucky University}\\
{\small\it Highland Heights, Kentucky 41099, U.S.A.}\\
\end{center}

\begin{center}
{\bf Abstract}
\end{center}

In this paper we have studied two non-linear black holes in 2+1 dimensions. They are regular and could have two horizons for chosen values of the parameters in the theory. Thermodynamics of the two black holes are studied in the extended phase space where the  pressure $P = -\Lambda/8 \pi$.
In order to satisfy the Smarr formula and the first law of thermodynamics, a renormalization parameter is introduced. Hence there is an additional thermodynamical parameter  for the black hole. We have also studied the two black holes in the context of heat engines. A thermodynamical cycle, consisting of two isobaric and two isochoric is considered. Efficiency is computed by varying the non-linear parameter and the electric charge of the
black hole. It is observed that when the non-linear parameter and the charge increases, the efficiency increases.  When compared to the charged BTZ black hole, the efficiency for the rectangle cycle is smaller for the  regular black holes.

\hspace{0.7cm}

{\it Key words}: regular black hole, non-linear electrodynamics,  heat engine, anti-de Sitter space, efficiency


\section{Introduction}	

Black holes in 2+1 dimensions provide insights into physical phenomena of black holes in a simpler setting compared to its counterparts in 3+1 dimensions. The well known BTZ black hole in 2+1 dimensions has been immensely useful in understanding variety of issues with regard to black holes \cite{btz}. On the other hand, non-linear electrodynamics has attracted much interest since the Born-Infeld non-linear electrodynamics \cite{born}. Some black holes in 2+1 dimensions in non-linear electrodynamics coupled to gravity has been proposed \cite{cataldo1}-\cite{mazh3}. In this paper, the focus is on the 2+1-dimensional regular black holes, proposed by He and Ma in Ref.~\cite{He:2017ujy}.

Black holes in anti-de Sitter space have been center of attraction due to variety of interesting properties they hold. The seminal work by Hawking and Page demonstrated that there is a first order phase transition between the Schwarzschild anti-de Sitter black hole and the thermal AdS space \cite{hawking}. The discovery of  phase transitions similar to Van der Waals liquid/gas transitions in Reissner-Nordstrom AdS black holes by Chamblin et al.~\cite{cham1,cham2} was another landmark in black hole thermodynamics. When the negative cosmological constant is considered as the thermodynamical pressure as $ P = - \Lambda/ ( 8 \pi)$, the resulting first law of black hole has a  $VdP$ term: here the mass of the black hole is treated as the enthalpy of the black hole as first presented by Kastor et al.~\cite{kastor}. One of the first works to explore  thermodynamics in AdS black holes where the negative cosmological constant is considered as the dynamical pressure was the paper by Kubiznak and Mann~\cite{mann}. There the thermodynamics of the charged black hole in 4 dimensions was studied in detail. Extended phase space in anti-de Sitter black holes have been studied extensively. Many black holes have demonstrated first order phase transitions similar to van de Waals phase transitions between liquid and gas.  There are large number of papers to mention on this topic: few examples are \cite{fernando1}-\cite{mann5}. Due to the limitation of space, we would instead mention an interesting review on {\it Black hole chemistry} by Kubiznak et.al. which has a comprehensive summary on thermodynamics of black holes in AdS space \cite{kubi}.

Another interesting property of AdS black holes is that they can be  used as heat engines. Since the pressure and the volume of the black hole is dynamic, one can extract mechanical work via the $PdV$ term. A suitable thermodynamical cycle can be defined so that there is a net work done by the system. In classical thermodynamics, there are several thermodynamical cycles that are used in heat engines: Carnot cycle, Otto cycle and Brayton cycle are few examples. Carnot cycle has the highest efficiency and is made of two pairs of isothermal and adiabatic processes. There have been several papers on AdS black holes as heat engines \cite{johnson1}-\cite{johnson4}. In this paper, we will present the non-linear black hole as a heat engine.

The paper is organized as follows: in section 2, the black hole considered is presented. In section 3 thermodynamics of the black holes are given. In section 4 the black holes are studied in the context of heat engines and finally in section 5 the conclusion is given.


\section{Regular 2+1-dimensional black holes}

A class of regular black holes with non-linear electrodynamic sources was proposed by He and Ma in Ref.~\cite{He:2017ujy}. The corresponding action is given by,
\begin{equation} \label{action}
S = \int d^4x \sqrt{-g} \left[ \frac{(R - 2 \Lambda)}{16 \pi G} + L(F) \right]
\end{equation}
Here, $L(F)$ is given by the Lagrangian for the non-linear electrodynamics were $F = F_{ \mu \nu} F^{ \mu \nu}$, $g$ is the determinant of the metric tensor and $\Lambda = -\frac{1}{l^2}$ is the Cosmological constant. The metric was given in the form
\begin{equation}
ds^2 = -f(r)dt^2 + \frac{ dr^2}{f(r)} + r^2 d \phi^2
\end{equation}
The electromagnetic tensor is $F_{\mu \nu} = E(r) \left( \delta ^t_{\mu} \delta^r_{\nu} - \delta^t_{\nu} \delta^r_{\mu} \right)$, where $E(r)$ is the electric field.

In this paper, we will consider two static regular black hole solutions presented  in Ref.~\cite{He:2017ujy} by He and Ma. They are referred to as case II and case III in the paper and we will follow the same name for the two black holes.


The case II solution is defined by the metric function
\begin{equation}
f(r) = - m + \frac{r^2}{l^2} - q^2 \ln\left(\frac{q^2}{a^2 l^2} + \frac{r^2}{l^2}\right)
\,\,\label{reg-II} \, 
\end{equation}
where $l$ is related to the cosmological constant by $\Lambda = -1/l^2$ and the corresponding electric field is given by
\begin{equation}
E(r) = \frac{a^4 q \, r^3}{16 \pi \left(q^2 + a^2 r^2 \right)^2}
\,\,\label{E-reg-II} \, 
\end{equation}
Black hole mass $M = \frac{m}{8}$ and the electric charge $Q = 8 \pi q$.
This black hole solution was also derived by Cataldo and Garcia~\cite{Cataldo:2000ns} with two differences: instead of $- q^2 \ln\left(\frac{q^2}{a^2 l^2} + \frac{r^2}{l^2}\right)$ term in the metric function $f(r)$, they had $- q^2 \ln\left(a^2  + r^2\right)$. The black hole in case II has two, one or no horizons for positive mass.


On the other hand the case III solution is defined by the following metric function
\begin{equation}
f(r) = - m + \frac{r^2}{l^2} - 2 q^2 \left[\ln\left(\frac{q}{al} + \frac{r}{l}\right) +\frac{q}{a r + q} \right]
\,\,\label{reg-III} \, 
\end{equation}
and the corresponding electric field is given by
\begin{equation}
E(r) = \frac{a^3 q r^2}{16 \pi (a r+q)^3}
\,\,\label{E-reg-III} \, 
\end{equation}
Black hole mass $M = \frac{m}{8}$ and the electric charge $Q = 8 \pi q$. For case III, there is a horizon when $ 0 < a < e^{3/2}$ and two horizons when $ a > e^{3/2}$.

In both case I and case II, the electric field asymptotically behaves as the 2+1-dimensional Maxwell solution and it vanishes as $r \rightarrow 0$.

Note that with these solutions we recover the charged BTZ black hole~\cite{Martinez:1999qi} when $a \rightarrow \infty$. Similarly when $r \rightarrow \infty$ this solution behaves as the charged BTZ black hole.


\section{Thermodynamics of black holes in 2+1 dimensions}
Before we proceed to discuss thermodynamics of the non-linear black holes in section~2, some discussion of 2+1-dimensional charged black hole in Maxwell electrodynamics (well known as charged BTZ black hole) is in order. The charged BTZ black hole is given with the metric function 
\begin{equation}
f(r) = - m -  2 q^2 \ln\left(\frac{r}{l}\right) + \frac{ r^2}{l^2}
\end{equation}
In 2+1 dimensions, the Smarr formula~\cite{Smarr:1972kt} for a static charged black hole is given by $T S - 2 P V = 0$, as demonstrated by Frassino et.al. in Ref.~\cite{Frassino:2015oca}. The first law of thermodynamics of black holes with pressure defined as $ P = - \frac{\Lambda}{8 \pi} = \frac{1}{ 8 \pi l^2}$ and mass M (considered as the enthalpy) is given by the expression $dM = T dS + V dP + \Phi dQ$. Here, entropy $S = \pi r_+/2$, $T = f'(r_+)/ 4 \pi$ and $\Phi$ is the potential conjugate to the electric charge $Q$. It was shown in Ref.~\cite{Frassino:2015oca} that the Smarr formula and the first law of thermodynamics  for the charged BTZ black hole is satisfied if the relevant thermodynamic quantities are defined as,
\begin{equation}
 V =  \left(\frac{\partial M}{\partial P}\right)_{S,Q} =  \pi r_+^2 - q^2 \pi l^2
\end{equation}
\begin{equation}
\Phi = \left(\frac{\partial M}{\partial Q}\right)_{S,P} =   - \frac{q}{16 \pi} \ln\left( \frac{r_+}{l}\right)
\end{equation}
Here  the mass is defined as
\begin{equation}
M = \frac{m}{8} = \frac{r_+^2}{ 8 l^2} - \frac{q^2}{4} \ln \left( \frac{r_+}{l}\right)
\end{equation}
In this approach, the volume depends on the charge $Q$ and the Reverse Isoperimetric Inequality is violated \cite{Frassino:2015oca}. 

An alternate approach was presented in Ref.~\cite{Frassino:2015oca}, where $V = \pi r_+^2$: a renormalization length scale $R$ is introduced where the metric function is rewritten as 
\begin{equation}
f(r) = - m_0 + \frac{r^2}{l^2} -  2 q^2 \ln \left( \frac{r}{R}\right)
\end{equation}
Here $m_0 = m -  2 q^2 \ln(l/R)$. Now the mass is, 
\begin{equation}
M = \frac{m_0}{8} = \frac{r_+^2}{ 8 l^2} - \frac{q^2}{4} \ln\left( \frac{r_+}{R}\right)
\end{equation}
A new thermodynamic variable, 
$K =  \left(\frac{\partial M}{\partial R}\right)_{S,Q,P}$ is introduced as the conjugate to the parameter $R$. The modified Smarr formula and the first law of thermodynamics are,
\begin{equation}
TS - 2 P V + K R =0 \,\,\,\,\,  ;  \,\,\,\,\,  dM = T dS + V dP + \Phi d Q + K dR
\,\,\label{two-eqs} \,
\end{equation}
So that in the limit $r\rightarrow \infty$, then $R\rightarrow \infty$, thus allowing $r/R = 1$~\cite{Cadoni:2007ck}.

When regular black hole in case II is studied for its thermodynamic properties, in the extended phase space, the Smarr formula, $T S - 2 P V =0$ is satisfied. The first law of thermodynamics, $d M = T dS + V dP + \Phi dQ$ is also satisfied with the mass $M$ given as
\begin{equation}
M = \frac{r_+^2}{8 \, l^2}-\frac{q^2 }{8}  \ln
   \left(\frac{q^2}{a^2}+\frac{r_+^2}{l^2}\right)
\end{equation}
However, in order to achieve this, a slight modification of the metric function has to occur as \cite{He:2017ujy},
\begin{equation}
f(r) = - m + \frac{r^2}{l^2} - q^2 \ln\left(\frac{q^2}{a^2} + \frac{r^2}{l^2}\right)
\end{equation}
In this case the volume is given by
\begin{equation}
V = \left(\frac{\partial M}{\partial P}\right)_{S,Q} =  \pi  r_+^2 \left(1-\frac{a^2 l^2 q^2}{a^2 r_+^2+l^2 q^2}\right)
\end{equation}
which depends on the charge $q$  and the Reverse Isoperimetric Inequality is violated just like for the charged BTZ black hole. Furthermore,  with this modification the parameter $l$ appears in the electric field. On the other hand, in the weak field approximation the problem associated with the term $\ln(r)$ is inherited.

\section{Thermodynamics of black hole in case II and case III}

In order to rectify the problems mentioned, and to obtain volume $V = \pi r_+^2$ while satisfying the Smarr formula  and the first law of thermodynamics, we will use the second approach presented in Ref.~\cite{Frassino:2015oca}:  we introduce a new thermodynamic parameter $R$ associated with the renormalization length scale. The non-linear parameter $a$ is redefined as $\frac{a}{R}$ in both metrics where $R$ is a renormalization length scale.

After rescaling $a$ as $a/R$, the metric in case II solution simplifies to
\begin{equation}
f(r) = - m + \frac{r^2}{l^2} - q^2 \ln \left(\frac{q^2 R^2}{l^2 a^2} + \frac{r^2}{l^2}\right)
\,\,\label{reg-II-R} \,  
\end{equation}
which is rewritten as,
\begin{equation}
f(r) = - m_0 + \frac{r^2}{l^2} - q^2 \ln\left(\frac{q^2}{a^2} + \frac{r^2}{R^2}\right)+ q^2 \ln\left(\frac{q^2}{a^2} + 1 \right)
\,\,\label{reg-II-m0} \, 
\end{equation}
where 
\begin{equation}
m_0 = m + q^2 \ln\left(\frac{R^2}{l^2}\right) + q^2 \ln\left(\frac{q^2}{a^2} + 1 \right)
\,\,\label{m0-II} \, 
\end{equation}
And $f(r) \rightarrow  -m_0 +r^2/l^2$ as $r\rightarrow \infty$. 

The corresponding electric field is 
\begin{equation}
E(r) = \frac{a^4 q r^3}{ 16 \pi \left(a^2 r^2+q^2 R^2\right)^2}
\,\,\label{E-reg-II-R} \,  
\end{equation}
Note that the last term in the metric function does not contribute to it. In the limit $r \rightarrow \infty$,  $r/R \rightarrow 1$, so, $E(r)$ becomes
\begin{equation}
E(r) = \frac{q'}{16 \pi r}
\,\,\label{E-reg-II-R-asymp-eeff} \,  
\end{equation}
where this effective charge is $q' = q\left(1 + \frac{q^2}{a^2}\right)^{-2}$.
If $a \rightarrow \infty$ then $q' = q$. This is the charged BTZ black hole limit.
On the other hand, $E(r) \rightarrow 0$ as $r \rightarrow 0$.

If we define pressure as $P = - \frac{\Lambda}{ 8 \pi}$, then $l$ can be replaced as $l  = \sqrt{\frac{1}{ 8 \pi P}}$.
Hence the mass of the black hole $M$ in terms of $P$ is given by,
\begin{equation}
M =  \pi P r_+^2-\frac{q^2}{8}  \ln
   \left(\frac{q^2}{a^2}+\frac{r_+^2}{R^2}\right)+\frac{q^2}{8}  \ln
   \left(\frac{q^2}{a^2}+1\right)
 \,\,\label{M-K-II} \, 
 \end{equation}
here $r_+$ is the horizon radius. 
The Hawking temperature is given by
\begin{equation} \label{temp2}
T =  \frac{ 1}{ 4 \pi}  \left| \frac{ df(r)}{ dr} \right|_ { r = r_+} =  \frac{r_+}{2\pi } \left( 8 \pi P-\frac{a^2 q^2}{a^2 r_+^2+q^2 R^2}\right)
\end{equation}
The entropy of the black hole, $S = \pi r_+/2$. In order to verify the Smarr formula and the first law of thermodynamics,  we will compute the volume V, the electric potential $\Phi$ and  the thermodynamic conjugate to $R$, $K$ as follows 
\begin{equation}
V = \left(\frac{\partial M}{\partial P}\right)_{S,Q,R} = \pi r_+^2
\,\,\label{vol-2} \, 
\end{equation}
\begin{equation}
\Phi  =  \left(\frac{\partial M}{\partial Q}\right)_{S,P,R} = \frac{q}{32 \pi} \ln
   \left(\frac{R^2 q^2 + R^2 a^2}{R^2 q^2 + r_+^2 a^2}\right) +
   \frac{ q^3 a^2 ( r_+^2 - R^2)}{ 32 \pi  ( a^2 + q^2) ( a^2 r_+^2 + q^2 R^2)} 
\end{equation}
\begin{equation}
K = \left(\frac{\partial M}{\partial R}\right)_{S,Q,P} = \frac{q^2 r_+^2}{4 R^3 \left(\frac{q^2}{a^2}+\frac{r_+^2}{R^2}\right)}
\,\,\label{K-function-II} 
\end{equation}
With the above quantities, it can be shown that the Smarr formula and first law of the form (\ref{two-eqs})   are held.

The equation of state for the black hole can be described from Eq.~(\ref{temp2}) as
\begin{equation}
P= \frac{ T \sqrt{\pi}} { 4 \sqrt{V}} + \frac{ a^2 q^2}{ 8  ( a^2 V + \pi q^2 R^2)}
\end{equation}
It is observed that when $P$ vs $V$ is plotted, no criticality emerges, leading to no phase transitions for this black hole.


Now we follow the same procedure as was done for Case II in rescaling $a$ and reformulate $f(r)$ as follows
\begin{equation}
f(r) = - m_0 + \frac{r^2}{l^2} - 2 q^2 \left[\ln\left(\frac{q}{a} + \frac{r}{R}\right) +\frac{q R}{a r + q R} \right] +  2 q^2 \left[\ln\left(\frac{q}{a} + 1\right) +\frac{q}{a + q} \right]
\,\,\label{reg-III-m0} \,  
\end{equation}
and thus asymptotically behaves as $f(r) = - m_0 + \frac{r^2}{l^2}$, where 
\begin{equation}
m_0 = m - q^2 \ln \left(\frac{R^2}{l^2}\right)+ 2 q^2 \left[\ln\left(\frac{q}{a} + 1\right) +\frac{q}{a + q} \right]
\,\,\label{m0-III} \, 
\end{equation}

The corresponding electric field is given by
\begin{equation}
E(r) = \frac{a^3 q r^2}{16 \pi (a r+q R)^3}
\,\,\label{E-reg-III-R} \, 
\end{equation}
In  the limit $r \rightarrow \infty$ and maintaining  $r/R \rightarrow 1$, this solution behaves as
\begin{equation}
E(r) = \frac{q'}{16 \pi r}
\,\,\label{E-reg-III-R-asymp-eeff} \, 
\end{equation}
where the effective charge $q'$ is $q' = q\left(1 + \frac{q}{a}\right)^{-3}$.

And, as above, $l$ can be replaced in terms of $P$ and the mass $M$ of the black hole can be written as
\begin{equation}
M =  \pi r_+^2 P -  \frac{q^2}{4} \left[\ln\left(\frac{q}{a} + \frac{r_+}{R}\right) +\frac{q R}{a r_+ + q R} \right] +  \frac{q^2}{4} \left[\ln\left(\frac{q}{a} + 1\right) +\frac{q}{a + q} \right]
 \,\,\label{M-K-III} \, 
\end{equation}
The Hawking temperature is given by,
\begin{equation} \label{temp3}
T =  \frac{ 1}{ 4 \pi}  \left| \frac{ df(r)}{ dr} \right|_ { r = r_+} =  4  P r_+ -\frac{a^2 q^2 r}{2 \pi (a r_+^2+q R)^2}
\end{equation}
In order to verify the Smarr formula and the first law of thermodynamics, we compute the electric potential $\Phi$ and the thermodynamic conjugate to $R$, $K$ as follows
\begin{equation}
\Phi =  \left(\frac{\partial M}{\partial Q}\right)_{S,P,R}= \frac{ a q^2 (r_+ -R)( 4 a^2 r_+ + 2 q^2 R + 3 a q (r_++ R))}{ 32 \pi  ( a + q)^2 ( a r_+ + q R)^2} +
\frac{q}{ 16 \pi } \ln \left( \frac{ R a + R q}{ R q + a r_+} \right) 
\end{equation}
\begin{equation}
K = \left(\frac{\partial M}{\partial R}\right)_{S,Q,P} = \frac{ a^2 q^2 r_+^2}{ 4 R( a r_+ + q R)^2}
\end{equation}
The volume is similar to that of Eq.~(\ref{vol-2}).
With these quantities, it can be shown that the Smarr formula and first law of thermodynamics in the form given in Eq.(\ref{two-eqs}) are held.
The equation of state for the black hole can be described from Eq.~(\ref{temp3}) as
\begin{equation}
P = \frac{ T \sqrt{\pi}} { 4 \sqrt{V}} + \frac{a^2 q^2}{ 8  ( a V + \sqrt{\pi} q R)^2}
\end{equation}
When $P$ vs $V$ is plotted, no criticality emerges, leading to no phase transitions for this black hole.


\section{Black holes as heat engines}

\begin{figure} [ph]
\begin{center}
\includegraphics[width=0.4\textwidth]{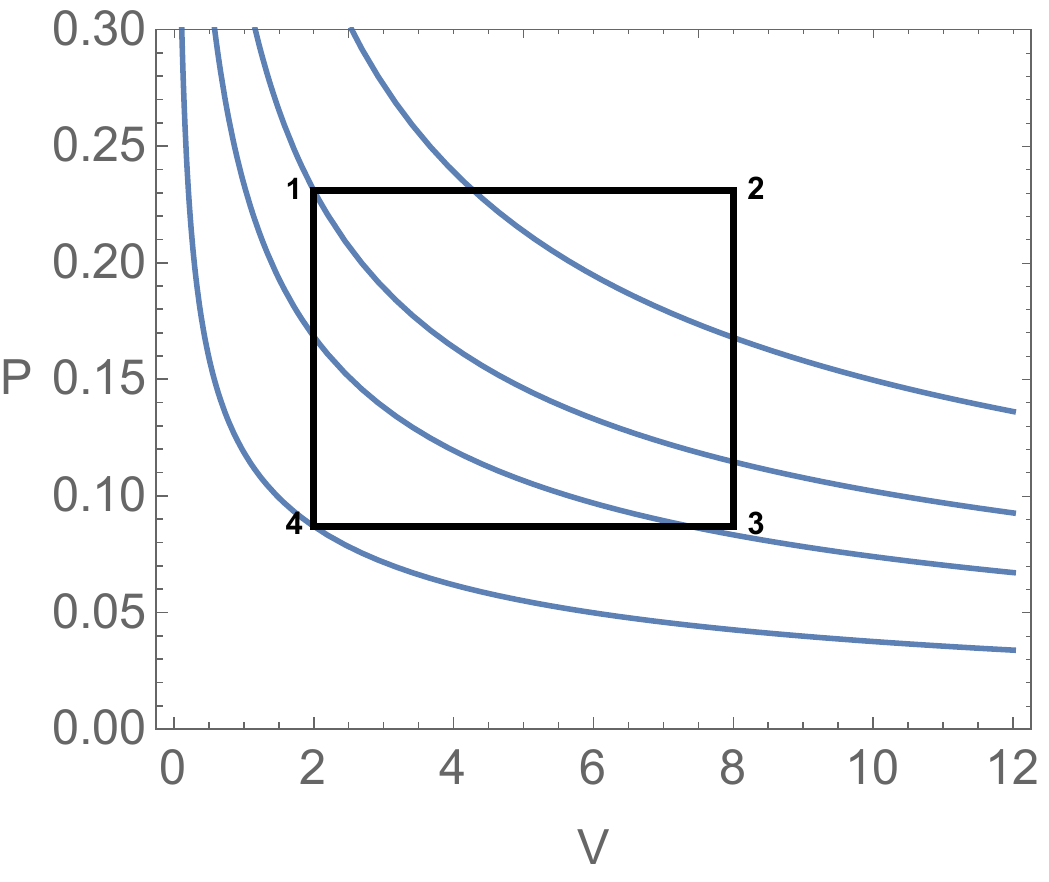}
\caption{The figure shows  $P$ vs $V$ for Case II for varying temperature. Here  $q = 1, R= 1, a= 1$. For large temperature the pressure is higher.}
\label{pvsv2}
 \end{center}
 \end{figure}

Now that an equation of state is clearly defined for the above black holes, one could use them as heat engines to do useful work. This is a special feature of black holes in extended phase space with a pressure term. The $PdV$ term in the first law provides the mechanical work of the heat engine and the working substance is the black hole solution discussed earlier in the paper. For a heat engine, the black hole is set in a thermodynamical cycle and let it produce work via the $PdV$ term. Heat will flow into the system and out of the system in certain part of the cycle. The efficiency of the heat engine is calculated by $\eta = \frac{ W_{net}}{Q_H}$ where $W_{net}$ is the net work done by the black hole and $Q_H$ the net heat flow into the black hole. If $Q_C$ is the net heat flow out of the black hole, $W_{net} = Q_H - Q_C$.


In this paper, we define a closed cycle in the state space with  two isochoric and two isobaric  paths as given in Fig.~\ref{pvsv2}. The efficiency of the cycle is given by
\begin{equation}
\eta = 1-\frac{M_3 - M_4}{M_2 - M_1}
\,\,\label{effi-gral} \,  
\end{equation}
The above formula was obtained by using the enthalpy and first law of thermodynamics by Johnson~\cite{Johnson:2016pfa} for the specific cycle in Fig.~\ref{pvsv2}.   

Since the entropy of a black hole is proportional to the volume of it, the isochoric path and the adiabatic paths coincide for a thermodynamic cycle for a black hole.


\begin{figure} [H]
\begin{center}
\includegraphics[width=0.9\textwidth]{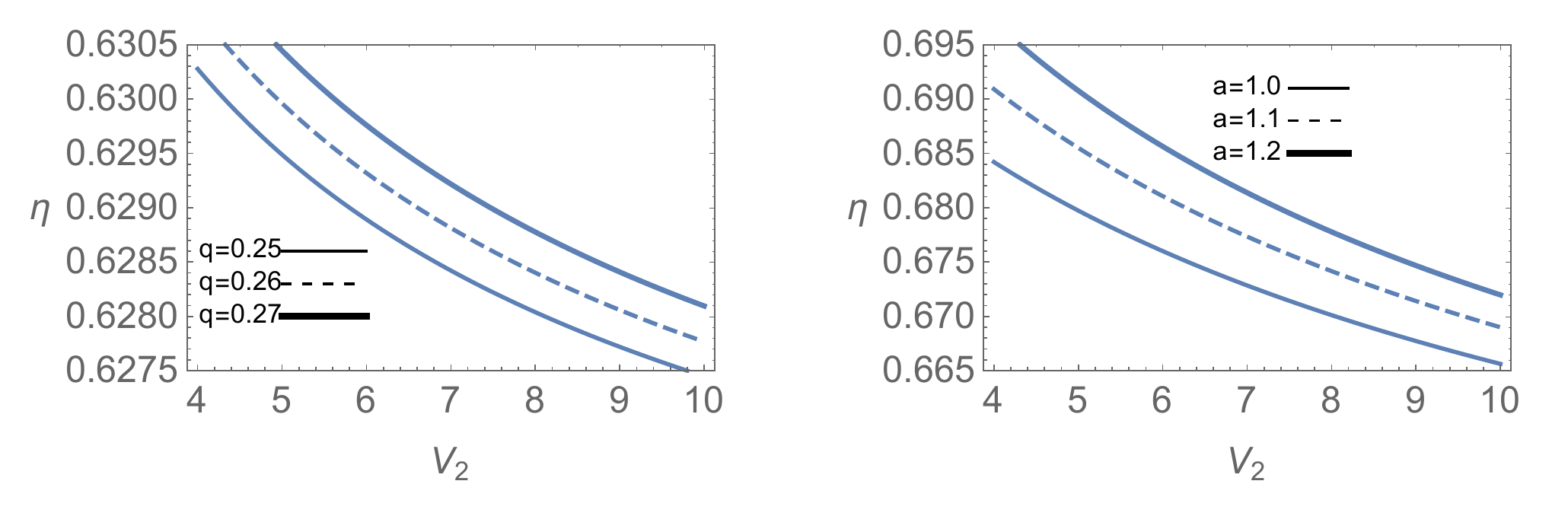}
\caption{The figure shows  $\eta$ vs $V_2$ for case II for varying $q$ and $a$ respectively. In both cases $R$ is fixed at $1$.When $q$ is varied, $a =1$ and when $a$ is varied, $q =1$}
\label{effi2}
 \end{center}
 \end{figure}

Now we will compute the efficiency for the cycle given above. There are two parts for the computation: in both cases $R$ is held fixed. First, the efficiency is computed by varying the charge $q$ and keeping $a$ fixed. Second, the efficiency is computed by varying the non-linear parameter $a$ while $q$ is fixed. In both computations, the volume $V$ and  $P$ for the coordinates 1 and 4 are kept constant. Here the volume $V_2 = V_3$  is varied to see how the efficiency changes with it (Fig.~\ref{effi3}). For chosen values for the parameters, the mass of the black holes are computed to make sure that it is positive. 


\begin{figure} [ph]
\begin{center}
\includegraphics[width=0.5\textwidth]{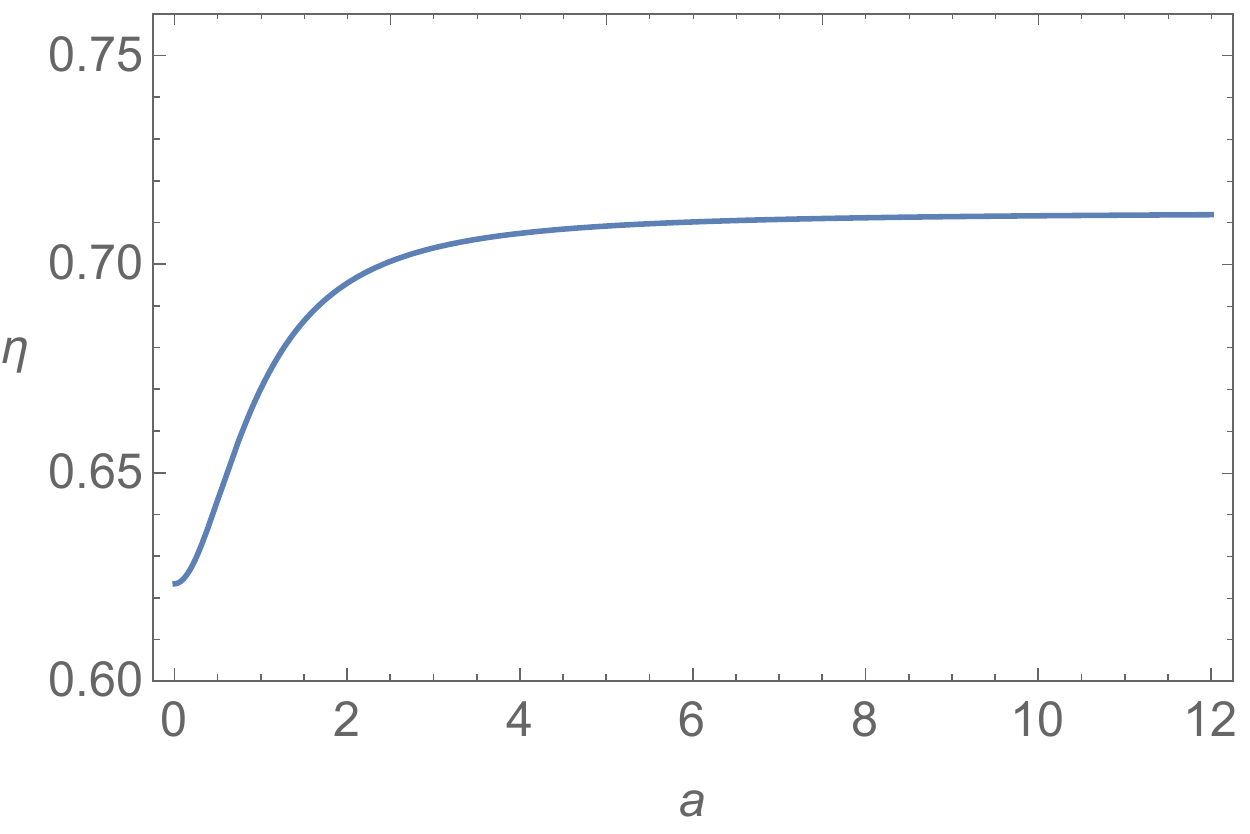}
\caption{The figure shows  $\eta$ vs $a$ for case II. Here  $q$ and $R$ are held fixed: $q = 1$ and $R= 1$.}
\label{effia2}
 \end{center}
 \end{figure}


\begin{figure} [ph]
\begin{center}
\includegraphics[width=0.9\textwidth]{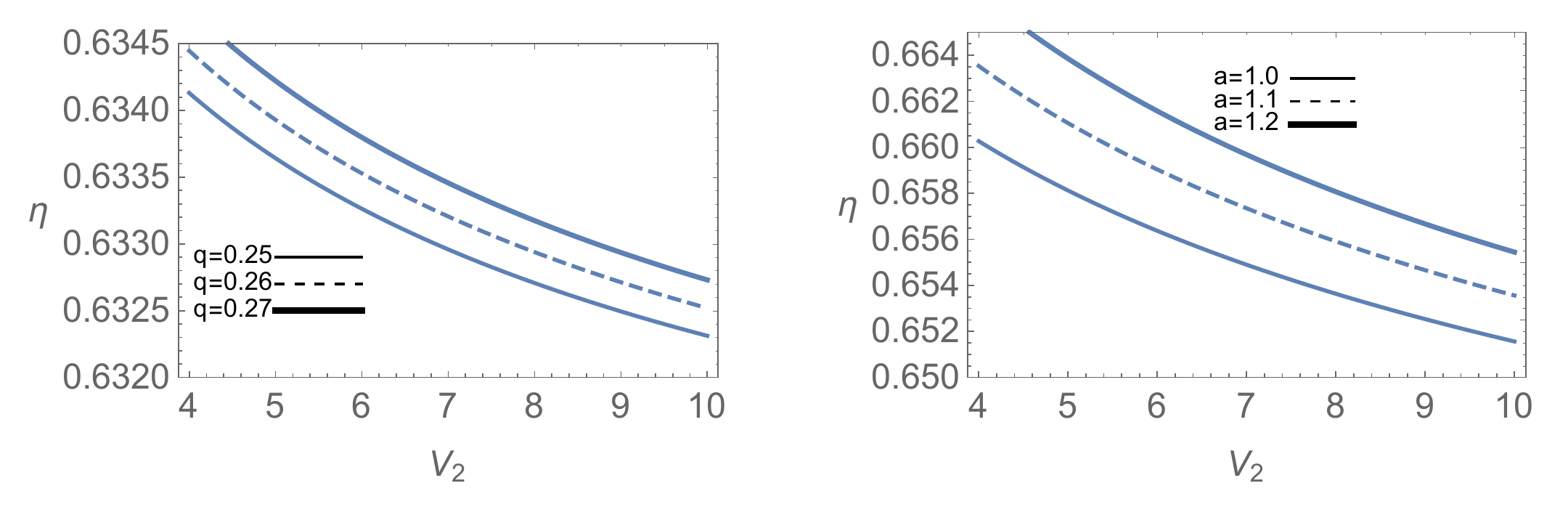}
\caption{The figure shows  $\eta$ vs $V_2$ for case III for varying $q$ and $a$ respectively. In both cases $R$ is held fixed at $1$. When $q$ is varied, $a =1$ and when $a$ is varied, $q =1$}
\label{effi3}
 \end{center}
 \end{figure}

\begin{figure} [ph]
\begin{center}
\includegraphics[width=0.5\textwidth]{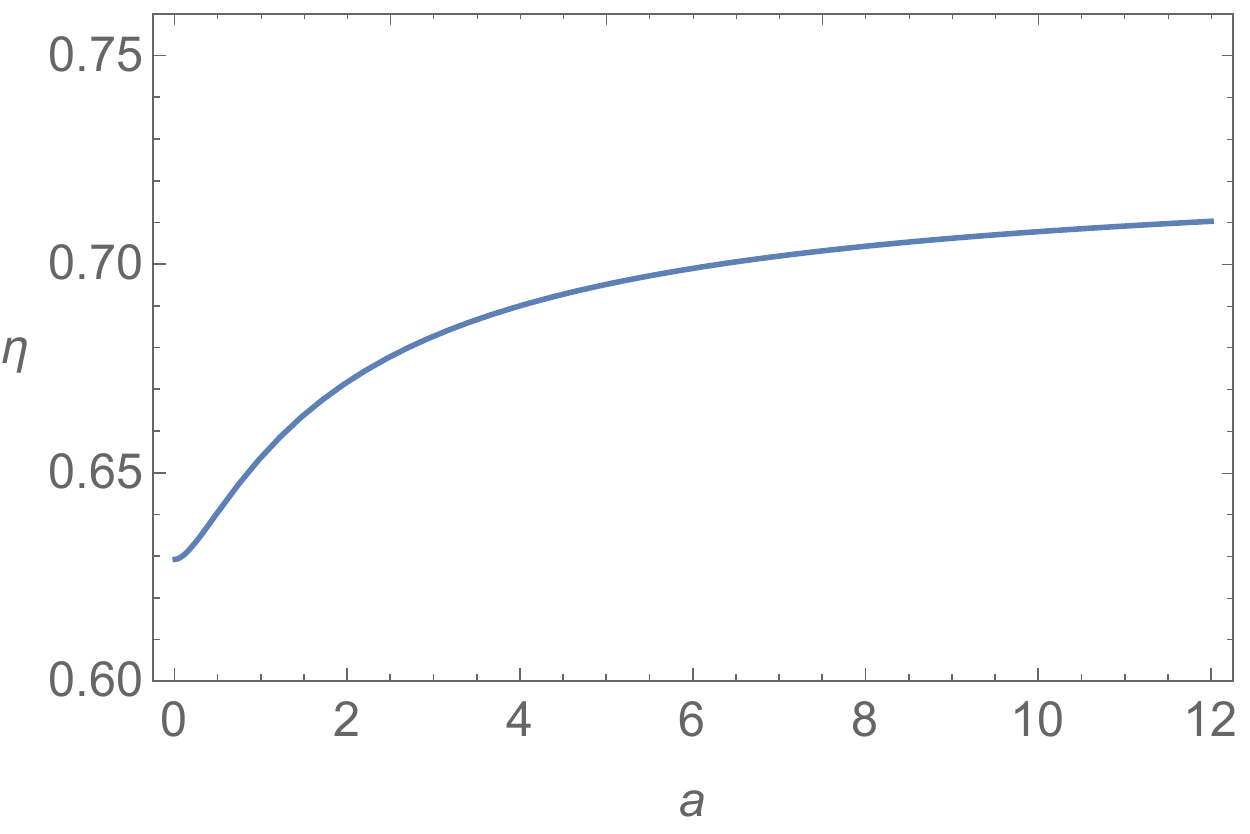}
\caption{The figure shows  $\eta$ vs $a$ for case III. Here  $q$ and $R$ are held fixed: $q = 1$ and $R= 1$.}
\label{effia3}
 \end{center}
 \end{figure}

Utilizing Eq.~(\ref{M-K-II}) and definitions of $P$ and $S$, the efficiency for case II is obtained as 
\begin{equation}
\eta = \frac{32 \left(P_1-P_4\right) \left(S_2^2-S_1^2\right)}{32 P_1
   \left(S_2^2-S_1^2\right) -  \pi  q^2 \ln F(S_1,S_2)}
\,\,\label{effi-regular-R-II} \,  
\end{equation}
where, for short, we have introduced the following definition
\begin{equation}
F(S_1,S_2) = \frac{4 a^2 S_2^2+\pi ^2 q^2 R^2}{4 a^2
   S_1^2+\pi ^2 q^2 R^2}
\,\,\label{F-def} \,  
\end{equation}

From Eq.~(\ref{M-K-III}) the efficiency can be obtained as
\begin{equation}
\eta = \frac{16 \left(P_1-P_4\right) \left(S_2^2-S_1^2\right)}{16 P_1
   \left(S_2^2-S_1^2\right)-\pi  q^2 G\left(S_1,S_2\right)}
\,\,\label{effi-regular-R-III} \,  
\end{equation}
where we have introduced the following definition
\begin{equation}
G(S_1,S_2) = \ln
   \left(\frac{2 a S_2+\pi  q R}{2 a S_1+\pi  q R}\right) + \frac{1}{\frac{2 a S_2}{\pi  q R}+1} - \frac{1}{\frac{2 a S_1}{\pi  q R}+1}
\,\,\label{G-def} \,  
\end{equation}

Note that if $a \rightarrow \infty$ either in Eq.~(\ref{effi-regular-R-II}) or Eq.~(\ref{effi-regular-R-III}), then we recover the expression of the heat engine efficiency for the charged BTZ black hole~\cite{Mo:2017nhw}.

\section{Conclusions}

In this paper, we have studied two regular black holes in 2+1 dimensions arising in non-linear electrodynamics with a cosmological constant. Both  black holes could have two horizons for appropriate parameters of the theory. Both black holes are regular  at the origin and in the weak field limit they approximate to the Maxwell electrodynamics. 

Thermodynamics of the black holes are studied in the extended phase space where the pressure $P = -\Lambda/8 \pi$. In order for the black hole to satisfy the Smarr formula and the first law of thermodynamics, an approach presented in Ref.~\cite{Frassino:2015oca} is followed. In this approach, a new parameter, a renormalization length scale $R$  is introduced. It allows for a volume which is equal to the geometric volume of the black holes and avoids the violation of the Reverse Isoperimetric Inequality. As a result, there are two additional thermodynamic variables, $R$ and $K$. Here $K$ is the conjugate quantity to $R$. An equation of state is derived for both black holes. When the pressure $P$ against the  volume $V$ is plotted, the behavior is similar to an ideal gas. Hence there are no phase transitions for both black holes.

Since there is a valid first law of thermodynamics, the non-linear black holes can be treated as heat engines. The thermodynamical cycle considered in this work is a rectangle where there are two isochoric and isobaric processes. The efficiency of the heat engine is computed by varying the non-linear parameter $a$ and the charge $q$. It was observed that when the charge and the non-linear parameter increased, the efficiency increases for both black holes. In both case II and case III black holes, for $a \rightarrow \infty$, the black hole approaches the charged BTZ black hole. Hence from Fig. 3 and 5, it is clear that the efficiency increases and become a stable value. Hence it could be concluded that the charged BTZ black hole has a higher  efficiency than both non-linear black holes. To authors knowledge there are no studies done to compare the BTZ black hole heat efficiency with the efficiency of the charged BTZ black hole which may be an interesting avenue to do further study.

In extending this work, it would be interesting to study the black holes in 2+1 dimensions where the electric field is Coulomb like \cite{Cataldo:2000we}. In that case it may not be necessary to introduce the renormalization parameter $R$ since there would not be a logarithmic term in the metric.


\section*{Acknowledgments}

L.B. is supported by DIUFRO through the project DI19-0052.


\end{document}